\documentclass[aps,pre,letter,twocolumn,superscriptaddress,floatfix,longbibliography,reprint]{revtex4-2}

\usepackage{amsmath,amssymb} 
\usepackage{bm} 
\usepackage{graphicx} 
\usepackage{comment} 
\usepackage{textcomp} 

\usepackage{enumitem}
\setlist{noitemsep,leftmargin=*,topsep=0pt,parsep=0pt}

\usepackage{xcolor} 
\definecolor{lightgray}{gray}{0.6}
\definecolor{medgray}{gray}{0.4}

\usepackage{hyperref}
\hypersetup{
colorlinks=true,
urlcolor= blue,
citecolor=blue,
linkcolor= blue,
}

\newcommand{\mytitle}{Large deviations of a tracer position in the dense and the dilute limits of a single-file diffusion}

\begin{document}

\title{\mytitle}

\author{Jagannath Rana}
\email[]{jagannath.rana@tifr.res.in}
\author{Tridib Sadhu}
\affiliation{Department of Theoretical Physics, Tata Institute of Fundamental Research, 1 Homi Bhabha Road, Mumbai 400005, India}

\date{\today}

\begin{abstract}
We apply the macroscopic fluctuation theory to analyze the long-time statistics of the position of a tracer in the dense and the dilute limits of diffusive single-file systems. Our explicit results are about the corresponding large deviation functions for an initial step density profile with the fluctuating (annealed) and the fixed (quenched) initial conditions. These hydrodynamic results are applicable for a general single-file system and they confirm recent exact results obtained by microscopic solutions for specific model systems.
\end{abstract}

\maketitle
\begin{figure}
    \centering
    \includegraphics[width=8
    cm]{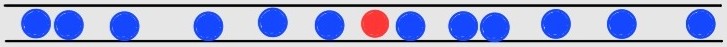}
    \caption{(Color online). A realization of a single-file system where particles confined in a narrow channel are constrained such that no particle can cross each other. The red particle denotes a tracer in the bath of identical other particles (blue).}
    \label{fig:Single_file}
\end{figure}

A one-dimensional interacting many-particle system with a restriction that particles cannot bypass each other is called a single-file system. Due to the single-file constraint, a large displacement of an individual particle needs to push surrounding particles in the same direction (see Fig.~\ref{fig:Single_file}). This caging effect leads to non-trivial transport properties. A tracer particle follows sub-diffusion with a diffusivity that is sensitive to the initial condition even at large times \cite{Leibovich2013,Krapivsky2015PRL,*Krapivsky2015tagged,Sadhu_2015}. In general, for large times, the sub-diffusion corresponds to the fractional Brownian motion with Hurst exponent $H=1/4$ \cite{Sadhu_2015,Leibovich2013,Krapivsky_2015_JSM}. 

The single-file system was introduced more than 60 years ago as a model to describe ion transport through cell membranes \cite{hodgkin1955potassium}. Since then a wide variety of physical, chemical and biological processes have been described using single file motion: molecular diffusion inside porous zeolite medium \cite{zeolite_1,Zeolite_2}, water transport inside carbon nanotube \cite{carbon_nanotube}, sliding of large protein molecules inside DNA \cite{dna_protein}, and transport of ions through super-ionic conductors \cite{super_ionic}, are just a few such examples.

The sub-diffusive nature was first theoretically shown by Harris \cite{harris1965diffusion} for Brownian point particles with hard-core repulsion and subsequently demonstrated in experimental systems \cite{sub_diff_1,sub_diff_2,sub_diff_3,sub_diff_4,carbon_nanotube}. There have been numerous attempts \cite{gen_file_1,gen_file_2,gen_file_4,Arratia1983} to extract the statistics of tracer position in general single-file systems with arbitrary interaction. Most general results are available for the mean and the variance of tracer position. Calculation of all the cumulants, equivalently the cumulant generating function (CGF) of tracer position in general single-file systems is still a challenging open problem.

A remarkable exact result for the tracer-CGF is in the single-file system of symmetric exclusion process where the result was derived by a solution of the microscopic dynamics for a fluctuating (annealed) initial state \cite{imamura2017large}. There are no analogous results available for general single-file systems. For a fixed (quenched) initial state of the symmetric exclusion process there are exact results for half-filling \cite{Sadhu_2015} and for the high density limit \cite{poncet2021cumulant}. Low density limit of the symmetric exclusion process corresponds to the hard-core Brownian particles which was exactly solved \cite{Krapivsky2015PRL,*Krapivsky2015tagged,gen_file_1}.

In this \textit{Letter} our main result is an exact expression for the CGF in a \textit{general} diffusive single-file system in its low and high density limits. Both annealed and quenched initial settings are considered. These results come as a straight-forward perturbation solution of a hydrodynamic approach presented earlier \cite{Krapivsky2015PRL,*Krapivsky2015tagged} and they confirm recent results \cite{illien2013active,imamura2017large,poncet2021cumulant} for specific model systems.

The hydrodynamic approach in \cite{Krapivsky2015PRL,*Krapivsky2015tagged} is an application of the  macroscopic fluctuation theory (MFT) \cite{mft_1,mft_2,mft_3,mft_4,mft_5} that was developed two decades ago and extends the Onsager-Machlup theory \cite{OnsagerMachlup1953} for far-from-equilibrium states. The theory is defined in a macroscopic scale where all microscopic details are embedded in a finite number of transport coefficients. For the diffusive single-file systems that we consider, the mobility $\sigma(\rho)$ and the diffusivity $D(\rho)$ are the relevant transport coefficients \cite{mft_5,Krapivsky2015PRL,*Krapivsky2015tagged} which are functions of the locally conserved macroscopic density of particles $\rho(x,t)$. In this approach the problem of calculating the CGF of tracer position in single-file reduces to a variational problem, which is hard to solve in general \cite{Krapivsky2015PRL,*Krapivsky2015tagged}. We show that the problem can be systematically approached using a perturbation expansion in density. To the leading order we get the CGF in the dilute and the dense limits for a general single-file diffusion. For the special case of symmetric exclusion process, our results confirm the expression of CGF obtained by microscopic solutions \cite{illien2013active,imamura2017large,poncet2021cumulant}. 

Besides generality, the hydrodynamic approach gives additional information about how the surrounding density profile is correlated with the tracer displacement, which are recently reported using microscopic calculations \cite{poncet2021generalised}.

\textit{Hydrodynamic formulation---}
In a coarse-grained description, time evolution of density field $\rho(x,t)$ in a single-file diffusion is given \cite{mft_5,Krapivsky2015PRL,*Krapivsky2015tagged} by the fluctuating hydrodynamics equation
\begin{equation}
    \partial_t\rho=-\partial_x j \quad \text{with}\; j= - D(\rho)\partial_x\rho+\sqrt{\sigma(\rho)}~\eta,
    \label{rho_dynamics}
\end{equation}
where $\eta(x,t)$ is a zero-mean Gaussian noise with covariance $\left<\eta(x,t)\eta(x',t')\right>=\delta(x-x')\delta(t-t')$. A local equilibrium condition relates the two transport coefficients to the free energy density $f(\rho)$ by a fluctuation-dissipation relation $2D(\rho)=\sigma(\rho)f''(\rho)$  \cite{derrida2007non}. Displacement of a tracer $X_t$ at time $t$ is related to the density field by the single-file constraint \cite{Krapivsky2015PRL,*Krapivsky2015tagged}
\begin{equation} 
    \int_{0}^{X_t}dx ~\rho(x,t)=
\int_{0}^{\infty}dx \,  \big( \rho(x,t)-\rho(x,0)\big),
	\label{def_X(rho)}
\end{equation}
where the tracer is assumed to be initially at the origin. This relation \eqref{def_X(rho)} gives the tracer position $X_t$ as a functional of the history of density $\rho(x,t)$. Different noise realizations of $\eta(x,t)$ in Eq.~\eqref{rho_dynamics} generate different histories for $\rho(x,t)$, which in turn results in different displacements of the tracer. Probability weight for a history of $\rho(x,t)$ is given by an Action, which comes straightforwardly following the Martin-Siggia-Rose-Janssen-de-Dominicis (MSRJD) formalism \cite{msrd_1,*msrd_@,Krapivsky2015PRL,*Krapivsky2015tagged,derrida2007non} for Eq.~\eqref{rho_dynamics}. Additional source of stochasticity comes from initial state. Considering probability of density fluctuations in the initial state $P(\rho(x,0))\sim e^{-F(\rho(x,0))}$ (the $\sim$ denotes leading dependence in the hydrodynamic scale), the generating function $\left<e^{\lambda X_T}\right>$ of the tracer position at time $T$ can be expressed as a path-integral \cite{Krapivsky2015PRL,*Krapivsky2015tagged}
\begin{equation}
    \left<e^{\lambda X_T}\right>=~\int~\mathcal{D}[\rho,\hat{\rho}]~ e^{-S_T[\hat{\rho},\rho]},
    \label{averaged_eqn}
\end{equation}
where $\hat{\rho}(x,t)$ is the MSRJD response field and the Action
\begin{align}
    S_T[\hat{\rho},\rho]&=\nonumber-\lambda X_T[\rho] + F[\rho(x,0)]+\int_{0}^{T}dt~\int_{-\infty}^{\infty}dx  \\ &\left(\hat{\rho}\;\partial_t \rho-\frac{\sigma(\rho)}{2}(\partial_x\hat{\rho})^2+D(\rho)(\partial_x\rho)(\partial_x \hat{\rho})\right).
    \label{optimized_action}
\end{align}
For the Action to be meaningful we assume that $\rho$ and $\hat{\rho}$ vanish at $x\to \pm \infty$, which does not affect the tracer statistics at finite $T$.

A re-scaling of the spatial coordinate by $\sqrt{T}$ and time by $T$, shows that $S_T$ is proportional to $\sqrt{T}$. Then for large $T$, the path integral in \eqref{averaged_eqn} is dominated by the path $(\hat{\rho},\rho)\equiv (p,q)$ that minimizes the Action and the cumulant generating function $\mu=\log \langle e^{\lambda X_T} \rangle$ of the tracer for large time $T$ is given by negative of the minimal Action. A variational calculation gives the least-Action path as a solution of \cite{Krapivsky2015PRL,*Krapivsky2015tagged}
\begin{equation}
    \begin{aligned}
    \partial_t p + D(q)\partial_{xx}p &= -\frac{\sigma'(q)}{2}(\partial_x p)^2\\
    \partial_t q - \partial_{x}\big(D(q)\partial_xq\big) &= -\partial_x\big(\sigma(q)\partial_x p\big).
    \end{aligned}
    \label{opti_field_equation}
\end{equation}
with appropriate boundary conditions that depend on the initial state.

To illustrate an unusual long-time memory effect for the single-file transport, two types of initial states are usually studied \cite{Leibovich2013,Krapivsky2015PRL,*Krapivsky2015tagged}. For the annealed case the initial state is in equilibrium, with \cite{derrida2007non,Krapivsky2015PRL,*Krapivsky2015tagged}
\begin{equation}
F(\rho(x))=\int_{-\infty}^{\infty}dx~\int_{\Bar{\rho}(x)}^{\rho(x)}dr~\frac{2D(r)}{\sigma(r)}(\rho(x)-r),\label{eq:F}
\end{equation}
where $\Bar{\rho}(x)$ is the mean-density profile of the initial state. In this case, the boundary conditions are on the field $p(x,t)$,
\begin{equation}
\begin{aligned}
    p(x,0)&=-\lambda \frac{\delta X_T}{\delta q(x,0)} + \frac{\delta F}{\delta q(x,0)},\\ p(x,T)&=\lambda \frac{\delta X_T}{\delta q(x,T)}.
\end{aligned}\label{eq:annealed bc}
\end{equation}
A quenched setting corresponds to the case where the initial density profile is fixed at the mean-density $\Bar{\rho}(x)$ and no fluctuations are allowed. In this case $F(\rho(x,0))=0$ and the boundary conditions are
\begin{equation}
    q(x,0)=\Bar{\rho}(x), \quad \text{and} \quad p(x,T)=\lambda \frac{\delta X_T}{\delta q(x,T)}.\label{eq:quenched bc}
\end{equation}

The names annealed and quenched are inspired by similar ensembles in disordered systems \cite{Mezard}. In analogy with the partition function in spin glass we define the annealed CGF $\mu_{\mathcal{A}}=\log\langle e^{\lambda X_T}\rangle_{\text{history+initial}}$ and the quenched CGF $\mu_{\mathcal{Q}}=\langle\log \langle e^{\lambda X_T}\rangle_{\text{history}} \rangle_{\text{initial}}$ where initial profile is analogous to disorder. In the latter definition, the logarithm being a slowly varying function compared to $e^{-F(\rho)}$, contribution to $\langle \rangle_{\text{initial}}$ is dominated by the mean-profile $\Bar{\rho}(x)$ and this justifies our choice for the fixed initial density in the variational formulation for the quenched case.

For both initial settings, the minimal Action reduces to a simple expression \cite{Krapivsky2015PRL,*Krapivsky2015tagged}
\begin{equation}
S_T[p,q]=-\lambda~Y+ F(q(x,0))+\int_{0}^{T}dt\int_{-\infty}^{\infty}dx~\frac{\sigma(q)}{2}(\partial_x p)^2,
\label{annealed_cgf_exp_1}
\end{equation}
with  $Y\equiv X_T[q]$ for the least Action path. Solution of the least-Action path \eqref{opti_field_equation} with appropriate boundary conditions and the appropriate $F[\rho]$-function gives the CGF $\mu\simeq -S_T[p,q]$ for large $T$ in the corresponding ensemble. 

An explicit solution for the least-Action path \eqref{opti_field_equation} for arbitrary $D(\rho)$ and $\sigma(\rho)$ is not available. A perturbation solution in $\lambda$ is possible \cite{Krapivsky2015PRL,*Krapivsky2015tagged} which leads to first few moments of tracer position, but not the CGF. We take an alternative avenue by treating density as perturbation parameter, which gives the CGF as a series in density. We consider two cases, the dilute and the dense limit, and determine the CGF for general single-file in both annealed and quenched settings.

Although the hydrodynamic formulation is applicable to arbitrary initial profile, we consider an example of step density profile $\Bar{\rho}(x)=\rho_a \Theta(-x)+\rho_b \Theta (x)$ which has been frequently studied for tracer diffusion \cite{illien2013active,imamura2017large,poncet2021cumulant,gen_file_3}. For $\rho_a \ne \rho_b$, bulk of the system evolves towards an asymptotic equilibrium effectively biasing the tracer in one direction. For the annealed setting, the two halves of the system are initially in equilibrium at different densities, and they are joined together for $t>0$.

\textit{Dense limit --}
The density in \eqref{rho_dynamics} is the dimensionless occupied-volume-fraction with a maximum value one. For simple exclusion process $\rho(x)dx$ gives the fraction of occupied sites in a hydrodynamic length between $x$ and $x+dx$. The limit where $\rho_a$ and $\rho_b$ are close to value one, there are very few number of vacant spaces in the single-file. In this limit the transport coefficients $D(\rho)\simeq  D(1)$, and $\sigma(\rho)\simeq (\rho-1)~ \sigma'(1)$. (Vanishing mobility for $\rho\to 1$ can be understood from the fluctuation-dissipation relation $2D(\rho)=\sigma(\rho)f''(\rho)$, and that in the dense limit leading contribution to the free energy density $f(\rho)$ comes from the positional entropy of voids.)

We consider an expansion for the least-Action path, $q=1+q_1+q_2+\cdots$ and $p=p_0 + p_1 + p_2 + \cdots$, where the subscript denotes the order in $1-\rho$. To leading non-trivial orders the least-Action path in Eq.~\eqref{opti_field_equation} follows
\begin{equation}
    \begin{aligned}
    D(1)^{-1}\partial_t p_0 + \partial_{xx}p_0 &= -\alpha(\partial_x p_0)^2,\\
    D(1)^{-1}\partial_t q_1 - \partial_{xx}q_1 &=-2 \alpha  \partial_x\big(q_1\partial_x p_0\big),
    \end{aligned}
    \label{step_1st_equation_1}
\end{equation}
where $\alpha=\sigma'(1)/2D(1)$. We shall see that solution at this order is sufficient for determining the leading term in the CGF.

A canonical transformation \cite{Derrida2009} $P=e^{\alpha p_0}$ and $Q=q_1e^{-\alpha p_0}$ reduces the equations \eqref{step_1st_equation_1} to a decoupled diffusion and an anti-diffusion equation which are easy to solve, leading to a general solution\begin{subequations}
\begin{align}
&e^{\alpha p_0(x,t)}=  \int_{-\infty}^{\infty}dz~e^{\alpha\, p_0(z,T)} ~g_{T-t}(z-x)\label{step_1_p_0}\\
&q_1(x,t)= \int_{-\infty}^{\infty}dz~q_1(z,0)~e^{-\alpha\big(p_0(z,0)-p_0(x,t)\big)}~g_t(z-x)
 \label{step_1_q_1}
\end{align}
with the diffusion kernel \label{eq:general solution}\end{subequations}
\begin{equation}
g_t(x)= \frac{\exp\left(-\frac{x^2}{4 D(1)t} \right)}{\sqrt{4\pi D(1)t}}.
\end{equation}
\textit{(a) Quenched case ---}
For the quenched case, perturbation expansion of the boundary condition in Eq.~\eqref{eq:quenched bc} gives $q_1(x,0)=\rho_a \Theta(-x)+\rho_b \Theta(x)-1$, and $p_0(x,T)=\lambda \Theta(x)$, where we used an expansion $X_T[q]\equiv Y=Y_1 + \cdots$. (Vanishing of $Y_0$ is understood from the absolute confinement of tracer in the fully packed limit.) Expanding Eq.~\eqref{def_X(rho)} we get 
\begin{equation}
    Y_1=\int_{0}^{\infty}dx  \big( q_1(x,T)-q_1(x,0)\big)\label{eq:y1 dense quench}
\end{equation}
and with this a perturbation expansion of Eq.~\eqref{annealed_cgf_exp_1} gives the leading order term of the CGF
\begin{equation}
    \mu_{\mathcal{Q}}(\lambda)\simeq \lambda Y_1 - \frac{\sigma'(1)}{2}\int_{-\infty}^{\infty}dx~q_1~(\partial_xp_0)^2
    \label{mu_Q}
\end{equation}
where we used $F[q]=0$.
The expression is further simplified \cite{suppl} to
\begin{align}
    \mu_{\mathcal{Q}}(\lambda)\simeq \nonumber -(1-\rho_b)&\int_{0}^{\infty}dx~\big(p_0(x,0)-\lambda\big)\\
    &-(1-\rho_a)\int_{0}^{\infty}dx~p_0(-x,0),
    \label{cumu_Qu}
\end{align}
by using \eqref{eq:y1 dense quench}, an identity
\begin{align}
    (1/&2)\sigma'(1)q_1 (\partial_x p_0)^2=\partial_t(q_1\,p_0)+\cr&\partial_x\left[D(1) (q_1\partial_x p_0-p_0 \partial_xq_1) + \sigma'(1) q_1 p_0 \partial_x p_0 \right]
    \label{identity}
\end{align}
that comes from Eq.~\eqref{step_1st_equation_1}, and by using the vanishing $q_1$ and $p_0$ at $x\to \pm \infty$.
An explicit expression for the CGF in Eq.~\eqref{cumu_Qu} is then obtained by using the solution for $p_0(x,0)$ in \eqref{step_1_p_0} for the quenched boundary condition, which gives,
\begin{subequations}
\begin{align}
\mu_Q(\lambda)\simeq  -\frac{\sqrt{4 D(1) T}}{\alpha}\; R_{\mathcal{Q}}(0,\alpha\lambda\vert 1-\rho_a,1-\rho_b)
 \label{quenched_cgf_form_1}
\end{align}
in the dense limit, where
\begin{align}
R_{\mathcal{Q}}(y,b & \vert r,s)
=r\int_{y}^{\infty}d\xi~\log\left[1+\frac{e^{b}-1}{2}\text{erfc}(\xi)\right]\cr 
& + s\int_{-y}^{\infty}d\xi~\log\left[1+\frac{e^{-b}-1}{2}\text{erfc}(\xi)\right].
 \label{eq:RQ}
\end{align}\label{eq:CGF Q dense final}
\end{subequations}

\textit{(b) Annealed case --}
The boundary condition \eqref{eq:annealed bc} for $p(x,T)$ is identical to that in the quenched case, therefore the solution for $p_0(x,t)$ from Eq.~\eqref{step_1_p_0} is same in both cases. A straightforward perturbation expansion of the second boundary condition in \eqref{eq:annealed bc} expresses $q_1(x,0)$ in terms of $p_0(x,0)$,
\begin{equation}
   q_1(x,0)=
    \begin{cases}
         -(1-\rho_a)e^{\alpha p_0(x,0)} ,\quad & \text{for } x\leq 0 \\
         -(1-\rho_b)e^{\alpha\big(p_0(x,0)-\lambda\big)} ,\quad &  \text{for } x > 0, \\
    \end{cases}
\end{equation}
which is then used in \eqref{step_1_q_1} for an explicit solution for $q_1(x,t)$.
Following a similar perturbation analysis of the minimal Action in Eq.~\eqref{annealed_cgf_exp_1}, and using the above boundary condition for $q_1(x,0)$ leads to a simple expression for the leading term of CGF in the dense limit,
\begin{align}
    \mu_{\mathcal{A}}(\lambda)&\simeq -\frac{(1-\rho_a)}{\alpha}~ \int_{-\infty}^{0}dx~\big(e^{\alpha p_0(x,0)}-1\big)\cr &-\frac{(1-\rho_b)}{\alpha}~ \int_{0}^{\infty}dx~\big(e^{-\alpha\lambda+\alpha p_0(x,0)}-1\big),
\end{align}
which with the solution for $p_0(x,0)$ gives an explicit expression for $\mu_{\mathcal{A}}(\lambda)$ that is almost identical in form to \eqref{quenched_cgf_form_1} except the $R_{\mathcal{Q}}$ replaced by $R_{\mathcal{A}}$ where
\begin{align}
    R_{\mathcal{A}}(y,b\vert r,s)&=r(e^{b}-1)\int_{y}^\infty d\xi~\frac{1}{2}\text{erfc}(\xi)  \nonumber\\
    &+s(e^{-b}-1)\int_{-y}^\infty d\xi~\frac{1}{2}\text{erfc}(\xi).
    \label{eq:RA}
\end{align}

\textit{Dilute limit ---}
The limit of small $\rho_a$ and $\rho_b$ corresponds to a very few number of particles compared to the space available. Intuitively, the dilute limit correspond to point particles with non-crossing condition where exact results are available \cite{Leibovich2013,Krapivsky2015PRL,*Krapivsky2015tagged,imamura2017large}. We confirm this intuition using perturbation solution of the general hydrodynamic theory in the low density limit. 

The analysis is similar to that is in the dense limit, and we present only the important steps. For simplicity we shall use similar notations used in the dense limit, but their meaning here will be restricted to the dilute limit unless mentioned otherwise. In the dilute limit, the diffusivity $D(\rho)\simeq D(0)$ and the mobility $\sigma(\rho)\simeq\rho \sigma'(0) $. (The vanishing mobility is by a similar reasoning as discussed in the dense limit.) Using an expansion of the least-Action paths, $q=q_1 + q_2 + \ldots$ and $p=p_0 + p_1 + p_2 + \ldots$ (the subscript denotes the order in density) in the Eqs.~\eqref{opti_field_equation} gives the equation followed by the leading non-vanishing terms $p_0$ and $q_1$ which are similar in form with Eq.~\eqref{step_1st_equation_1} except the difference that the terms are for the dilute limit, namely $\alpha=\sigma'(0)/2D(0)$ and $D(1)$ is replaced by $D(0)$. Their general solution is similar to \eqref{step_1_q_1}.

A crucial difference with the dense limit comes in the expansion $X_T[q]\equiv Y=Y_0+ Y_1 + \cdots $ in Eq.~\eqref{def_X(rho)}, where $Y_0$ is non-vanishing and given by the single-file constraint $\int_0^{Y_0}dx~q_1(x,T)=\int_{0}^{\infty}dx  \big( q_1(x,T)-q_1(x,0)\big)$. Intuitively this means that in the dilute limit the tracer can move (in contrast to the dense limit where $Y_0$ vanishes due to full packing). In fact, by dimensional argument, at low density, the tracer position is expected to scale with the inter-particle separation, and thereby inversely with density. This means the CGF in the dilute limit is expected to follow a scaling $\mu(\lambda)\simeq \rho\,h(\lambda/\rho)$. In our perturbation theory we take this into account by considering $\lambda$ of order of the density.

\textit{(a) Quenched case ---}
In the dense limit, the boundary condition \eqref{eq:quenched bc} gives a condition for the leading order $q_1(x,0)=\rho_a \Theta(-x)+\rho_b\Theta(x)$, and $p_0(x,T)=B\,\Theta(x-Y_0)$, where $B=\lambda/q_1(Y_0,T)$.
 
Similarly, expanding the minimal Action \eqref{annealed_cgf_exp_1} for the quenched case, and using the equation for $p_0$ and $q_1$ with their boundary conditions, we get \cite{suppl} the leading order term of the CGF in the dilute limit
\begin{align}
\mu_{\mathcal{Q}}(\lambda)\simeq \lambda~Y_0&-\int_{-\infty}^{\infty}dx~q_1(x,T)p_0(x,T)\cr
&+\int_{-\infty}^{\infty}dx~q_1(x,0)p_0(x,0),\label{eq:muQ light}
\end{align}
where $Y_0$ for the quenched case follows the single-file constraint $\int_{Y_0}^\infty dx~\big(q_1(x,T)-\rho_b\big)=Y_0\rho_b$.

The expression \eqref{eq:muQ light} requires the solution for $q_1(x,t)$ and $p_0(x,t)$ which is straightforward to get from the general solution \eqref{eq:general solution} by treating $B$ as a parameter in the boundary condition. The solution shows that $q_1(x,T)$ has a jump discontinuity at $x=Y_0$ and therefore $B$ can not be determined self-consistently from its definition. It needs to determined by further optimizing $\mu_\mathcal{Q}$ with respect to $B$ \cite{Krapivsky2015PRL,*Krapivsky2015tagged}.

\begin{subequations}
Using the explicit solution for $q_1(x,t)$ and $p_0(x,t)$ in \eqref{eq:muQ light} we get \cite{suppl} a parametric solution of the CGF in the dense limit 
\begin{equation}
   \mu_Q(\lambda)\simeq \frac{\sqrt{4 D(0) T}}{\alpha}\; \left\{\alpha\lambda y + R_{\mathcal{Q}}(y,b\vert \rho_a,\rho_b)\right\}
\end{equation}
with \eqref{eq:RQ}, where $y$ and $b$ are determined from
\begin{equation}
    \frac{\partial R_{\mathcal{Q}}}{\partial b}=0\quad \textrm{and}\quad\frac{\partial R_{\mathcal{Q}}}{\partial y}=-\alpha\lambda.\label{eq:CGF Q Dilute b}
\end{equation}
The two relations in \eqref{eq:CGF Q Dilute b} came respectively from the single-file condition for $Y_0$ and the additional optimization condition $d\mu_{\mathcal{Q}}/dB=0$.
\label{eq:CGF dilute quench}
\end{subequations}

\textit{(b) Annealed case ---} In the dilute limit, the boundary condition \eqref{eq:annealed bc} gives
\begin{equation}
    p_0(x,0)=B\Theta(x)+\frac{1}{\alpha}\log\frac{q_1(x,0)}{\Bar{\rho}(x)}
\end{equation}
and $p_0(x,T)=B\Theta(x-Y_0)$, where $B=\lambda/q_1(Y_0,T)$.  

Similar to the quenched setting, solution for $q_1(x,t)$ and $p_0(x,t)$ are determined by treating $B$ as a parameter which is to be determined from an optimization with respect to $B$.

To the leading order in the dilute limit, the minimal action in Eq. \eqref{annealed_cgf_exp_1} for the annealed case gives the CGF 
\begin{equation}
\mu_{\mathcal{A}}(\lambda)\simeq \lambda~Y_0+\frac{1}{\alpha}\int_{-\infty}^{\infty}dx~\big(q_1(x,0)-\Bar{\rho}(x)\big)
\label{mu_A_dilute}
\end{equation}
where we used a similar identity \eqref{identity} for dilute limit, the boundary condition for $p_0(x,T)$, and the single-file constraint $\int_{Y_0}^\infty dx \;q_1(x,T)=\int_0^\infty dx\; q_1(x,0)$ that is due to \eqref{def_X(rho)} in the dilute limit. The single-file condition gives $Y_0$ in terms of $B$ which is further determined from the optimization condition $d\mu_{\mathcal{A}}/dB=0$.

Incorporating solution for $q_1(x,t)$ in \eqref{mu_A_dilute} and the additional conditions for $Y_0$ and $B$, we obtain \cite{suppl} an expression for the annealed CGF in the dilute limit, that is almost identical in form to \eqref{eq:CGF dilute quench} except the $R_{\mathcal{Q}}$ replaced by $R_\mathcal{A}$ defined in \eqref{eq:RA}.

\textit{Large deviations:---}
Our primary results are the explicit expression for the CGF in two limiting densities for two different initial states. For the quenched case, the dense limit result is in \eqref{eq:CGF Q dense final} and the dilute limit result is in \eqref{eq:CGF dilute quench}. Their result for the annealed case are similar in form with $R_{\mathcal{Q}}$ replaced by $R_{\mathcal{A}}$ defined in \eqref{eq:RA}. These results are equivalent of the large time asymptotic of the probability of tracer position
\begin{equation}
    P\left(\frac{X_T}{\sqrt{4D_0 T}}=y\right)\sim \exp\left[-\frac{\sqrt{4D_0 T}}{|\alpha|}\;\phi(y)\right],
    \label{def_ldf}
\end{equation}
where $\phi(y)$ is the large deviation function (LDF) and $D_0$ is the leading diffusivity in the two density limits; $\alpha$ defined earlier in the two limits relates to the isothermal compressibility $K_T$ by $\alpha=\rho K_T$ in the dilute limit, and $\alpha=-K_T/(1-\rho)$ in the dense limit \cite{suppl}. The LDF relates to the CGF by a Legendre transformation and using the derived expression for the latter it is straightforward to obtain the following results:
In the dense limit,
\begin{align}
    \phi_{\mathcal{Q(A)}}(y)&\simeq -b y-R_{\mathcal{Q(A)}}(0,b \vert 1-\rho_a,1-\rho_b),\label{eq:ldf dense}
\end{align}
and in the dilute limit,
\begin{align}
    \phi_{\mathcal{Q(A)}}(y)&\simeq -R_{\mathcal{Q(A)}}(y,b\vert \rho_a,\rho_b),\label{eq:ldf dilute}
\end{align}
where for each case the parameter $b$ is determined by an optimization condition $\partial \phi/\partial b=0$.
\begin{figure}
    \includegraphics[width=0.48\textwidth]{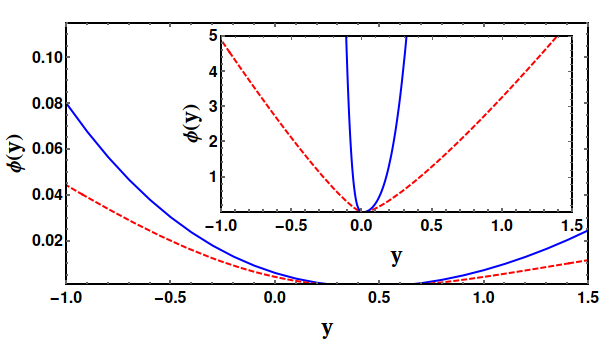}
    \caption{(Color online). The large deviation function $\phi(y)$ in \eqref{eq:ldf dilute} for the step initial density with $\rho_a=0.05$ and $\rho_b=0.01$ that corresponds to the dilute limit. Blue solid line denotes the quenched case and the red dashed line denotes the annealed case. The inset shows the corresponding results \eqref{eq:ldf dense} for the dense limit with $\rho_a=0.99$ and $\rho_b=0.95$. }
    \label{ldf_dense}
\end{figure}

Note that the LDF in \eqref{def_ldf} is independent of specific details of model systems which are in the parameters $D_0$ and $\alpha$. The results for $\phi(y)$ in Eqs.~\eqref{eq:ldf dense} match with the results \cite{illien2013active,poncet2021cumulant} in the dense limit of the symmetric exclusion process for which $D(\rho)=1$ and $\sigma=2\rho(1-\rho)$. Results in Eqs.~\eqref{eq:ldf dilute} agree with the results \cite{Krapivsky2015PRL,*Krapivsky2015tagged,Sadhu_2015} for hard-core Brownian point particles. 

In the annealed case, expression for LDF can be further simplified to an explicit formula. For the dilute limit we get
\begin{equation}
    \phi_{\mathcal{A}}(y)=\left(\sqrt{\rho_a h(y)}-\sqrt{\rho_b h(-y)} \right)^2
\end{equation}
with $h(y)=\frac{1}{2}\int_{y}^\infty \text{erfc}(x) dx$. (See \cite{suppl} for a similar formula in the dense limit.)

A comparative plot of the LDF for different cases is shown in Fig. \ref{ldf_dense} for step initial  profiles with $\rho_a>\rho_b$. The step initial state not only drifts the mean position of tracer, but it also makes fluctuations asymmetric around the mean as seen in the asymmetry of $\phi(y)$. Note that annealed LDF is wider than quenched LDF, which indicates a larger fluctuation for the former case. Similarly the narrower LDF for the dense limit (inset of Fig. \ref{ldf_dense}) compared to the dilute limit reflects that the tracer is less mobile in the former limit.

\textit{Concluding Remarks:---}
We presented analytical results for the long time statistics of the tracer position in a general single-file diffusion. Our results complement similar recent results \cite{illien2013active,poncet2021cumulant,Sadhu_2015,Leibovich2013} for specific model systems obtained using solution of microscopic dynamics. Our analysis presented here is a straightforward perturbation solution of the general theory reported earlier in \cite{Krapivsky2015PRL,*Krapivsky2015tagged}. The theory is based on a hydrodynamic formulation that, although less rigorous, gives the correct result for all cumulants at large times and it is applicable for a wider class of systems. Besides cumulants, the least action path $q(x,t)$ in our analysis gives how the density profile of surrounding particles evolve leading to a tracer position $X_T$.

In our perturbation approach, higher order terms could be systematically solved and that would give improved results for a wider range of density. It would be interesting to compare our general results for rare fluctuations in computer simulation of single-file with different inter-particle interactions. A particularly interesting case is when the tracer is confined in an external potential \cite{illien2013active,external_pot_1,external_pot_2}. Theoretically most challenging would be to extend the hydrodynamic approach for biased dynamics, when only the tracer is driven or when all particles are driven \cite{poncet2021cumulant,illien2013active,gen_file_3,Rajesh2001,Barkai2009,Majumdar1991,Demasi1985}. 

\begin{acknowledgments}
We acknowledge support of the Department of Atomic Energy, Government of India, under Project Identification No. RTI-4002. 
\end{acknowledgments}

\bibliography{references_LDF_Tracer_Extreme_Limits}

\end{document}


\title{ \Large{\textbf{Large deviation of a tracer position in the dense and the dilute \\ limits of a single-file diffusion}}\vspace{2mm}\\ {\large\textbf{\textit{{Supplementary Material}}}}\vspace{2mm}\\ \normalsize{Jagannath Rana and Tridib Sadhu}\\ \textit{\small{Department of Theoretical Physics, Tata Institute of Fundamental Research, 1 Homi Bhabha Road Mumbai 400005, India}}}

\maketitle
\begin{abstract}
We present here a few additional algebraic details for the calculation of the cumulant generating function (CGF) and the large deviation function (LDF) of position of a tracer, reported in the \textit{Letter}, in the dense and the dilute limits of general single-file diffusion. For ease of reading some algebraic details from the \text{Letter} are repeated. A derivation for the relation of $\alpha$, defined in the main text, with the isothermal compressibility $K_T$ is also presented.
\end{abstract}
\tableofcontents

\section{Variational formulation}
\paragraph{}
In a hydrodynamic description \cite{Krapivsky2015PRL,Krapivsky2015tagged}, the problem of the tracer-statistics in a general single-file diffusion reduces to a variational problem with an Action that is a functional of the density field $\rho(x,t)$ and the conjugate momentum field $\hat{\rho}(x,t)$. Corresponding minimal-Action gives the cumulant generating function of the tracer position. The optimal fields $(\rho,\hat{\rho})\equiv(q,p)$ for the minimal-Action are solution of a coupled set of differential equations (Eq.~(5) in the \textit{Letter}) with appropriate boundary conditions that depend on the initial state. They were first derived in \cite{Krapivsky2015PRL,Krapivsky2015tagged}. For convenience of the reader, we repeat some important details here.

For the boundary conditions on the optimal fields in the annealed case, computing the functional derivatives in Eq.~(7), using Eq.~(2) and Eq.~(6) of the \textit{Letter}, one gets  
\begin{equation}
    \begin{aligned}
    p(x,0)&=\frac{\lambda}{q(Y,T)}\Theta(x)+\int_{\Bar{\rho}(x)}^{q(x,0)}dz~\frac{2D(z)}{\sigma(z)}, \\\quad \text{and} \quad p(x,T)&=\frac{\lambda}{q(Y,T)}\Theta(x-Y),
    \end{aligned}
\label{eq:annealed bc}
\end{equation}
where we denote $X_T[q]\equiv Y$ where 
\begin{equation}
    \Bar{\rho}(x)=\rho_a \Theta(-x)+\rho_b \Theta(x)
\end{equation}
is the specific initial mean-density profile considered in our analysis. Similarly for the quenched case, corresponding boundary conditions are given in Eq.~(8) of the \textit{Letter}, which we repeat here,
\begin{equation}
    q(x,0)=\Bar{\rho}(x) \quad \text{and} \quad p(x,T)=\frac{\lambda}{q(Y,T)}\Theta(x-Y).
    \label{eq:quenched bc}
\end{equation}

The cumulant generating functions (CGF) in this hydrodynamic description are given by the negative of corresponding minimal Actions (Eq.~(9) of the \textit{Letter}), and expressed as
\begin{equation}
\mu_{\mathcal{A}}(\lambda)=\lambda~Y - F[q(x,0)]-\int_{0}^{T}dt\int_{-\infty}^{\infty}dx~\frac{\sigma(q)}{2}(\partial_x p)^2,
\label{an_cgf}
\end{equation}
and
\begin{equation}
\mu_{\mathcal{Q}}(\lambda)=\lambda~Y-\int_{0}^{T}dt\int_{-\infty}^{\infty}dx~\frac{\sigma(q)}{2}(\partial_x p)^2,
\label{qu_cgf}
\end{equation}
where the subscripts $\mathcal{A}$ and $\mathcal{Q}$ denote annealed and quenched respectively.

\section{Dense limit}  
\paragraph{}
The dense limit ($\rho_a,\rho_b\simeq 1$) corresponds to a very few number of vacancies in the system. The leading order optimal solutions are given in Eq.~(11) of the \textit{Letter}, which we use to calculate the CGFs in leading order. The difference in optimal paths, hence in CGFs, for the annealed and the quenched cases are due to their different boundary conditions in \eqref{eq:annealed bc} and \eqref{eq:quenched bc} respectively.
\subsection{Quenched case} 
\paragraph{}
We present the steps to get Eq.~(15) from Eq.~(14) of the \textit{Letter} and then steps to obtain Eq.~(17) of the \textit{Letter}. Inserting the identity of Eq.~(16) into Eq. (14) of the Letter, along with the vanishing boundary conditions of $q_1$ and $p_0$ at $x\to \pm \infty$, we arrive at,
\begin{equation}
\mu_{\mathcal{Q}}(\lambda)\simeq~ \lambda \int_{0}^{\infty}dx~q_1(x,T)-\lambda \int_{0}^{\infty}dx~  q_1(x,0)
+\int_{-\infty}^{\infty}dx~q_1(x,0)p_0(x,0)-\int_{-\infty}^{\infty}dx~q_1(x,T)p_0(x,T).
\label{mu_gen}
\end{equation}

Now plugging the boundary conditions for $q_1(x,0)$ and $p_0(x,T)$ (see \textit{Letter}), the first and the last term of the \eqref{mu_gen} get cancelled and we recover the Eq.~(15) of the \textit{Letter}.
The solution for $p_0(x,t)$ in Eq.~(11a) of the \textit{Letter} for the quenched boundary condition gives
\begin{equation} 
p_0(x,0)=\frac{1}{\alpha}\log\left[1+\frac{e^{\alpha\lambda}-1}{2}\text{erfc}\left(\frac{-x}{\sqrt{4D(1)T}}\right)\right].
\label{1}
\end{equation}
 
Note that using $\text {erfc}(-\xi)=2-\text {erfc}(\xi)$ one obtains,
\begin{align}
 p_0(x,0)-\lambda=~&\log\left[e^{-\alpha\lambda}+\frac{1-e^{-\alpha\lambda}}{2}\left\{2-\text{erfc}\left(\frac{x}{\sqrt{4D(1)T}}\right)\right\}\right]\cr
 =&\log\left[1+\frac{e^{-\alpha\lambda}-1}{2}\text{erfc}\left(\frac{x}{\sqrt{4D(1)T}}\right)\right].
 \label{2}
\end{align} 
Now the substitution of \eqref{1} and \eqref{2} along with the definition $\xi=\frac{x}{\sqrt{4D(1)T}}$ into Eq.~(15) in the \textit{Letter}, leads to the expression of quenched CGF reported in Eq. (17a) and (17b) in the \textit{Letter}.

\subsubsection*{Large deviation function}
\paragraph{}
The large deviation function of the tracer position (defined in Eq.~(25) in the \textit{Letter}) for the quenched case is related to the $\mu_\mathcal{Q}(\lambda)$ in Eq. (17a) of the \textit{Letter} by a Legendre transformation that gives 
\begin{equation}
    \phi_\mathcal{Q}(y)=-by-R_{\mathcal{Q}}(0,b) \qquad \text{with}\qquad y=-\partial_b R_{\mathcal{Q}}(0,b).
\end{equation}
(Note that here $\rho_a, \rho_b$ dependence is left implicit.) This is equivalent to the expression given in Eq. (26) in the \textit{Letter}.
\subsection{Annealed case}
\paragraph{}
Steps to obtain Eq.~(19) in the \textit{Letter} are not straightforward. These are presented here. Similar to the quenched case, we insert the optimal field expansions into Eq. \eqref{an_cgf}. But in this case we have an extra leading order term coming from the expansion of $F[q(x,0)]$ (in Eq.~(6) in the \textit{Letter}). So expanding $F[q(x,0)]$ we obtain in leading order,
\begin{align*}
    F&[q(x,0)]\simeq\frac{1}{\alpha}\int_{-\infty}^{\infty}dx~\int_{\Bar{\rho}(x)}^{1+q_1(x,0)}dz~\left(\frac{q_1(x,0)}{z-1}-1\right)\\
    &=\frac{1}{\alpha}\int_{-\infty}^{\infty}dx~\left[q_1(x,0)\log\Big[\frac{q_1(x,0)}{\rho_a\Theta(-x)+\rho_b\Theta(x)-1}\Big]-\Big\{1-\rho_a\Theta(-x)-\rho_b\Theta(x)+q_1(x,0)\Big\}\right].
\end{align*}

Now using Eq.~(18) of the \textit{Letter}, we finally obtain the CGF in the leading order,
\begin{align}
\mu_{\mathcal{A}}(\lambda)&\simeq\lambda~\int_{0}^{\infty}dx~\Big(q_1(x,T)-q_1(x,0)\Big)+\int_{-\infty}^{\infty}dx~q_1(x,0)p_0(x,0)-\int_{-\infty}^{\infty}dx~q_1(x,T)p_0(x,T)\cr
&~~~-\int_{-\infty}^{\infty}dx~q_1(x,0)~\Big\{p_0(x,0)-\lambda\Theta(x)\Big\} + \frac{1}{\alpha}\int_{-\infty}^{\infty}dx~\Big[1-\rho_a \Theta(-x) -\rho_b \Theta(x)+q_1(x,0)\Big] \cr
&=\frac{1}{\alpha}\int_{-\infty}^{\infty}dx~\Big[1-\rho_a \Theta(-x) -\rho_b \Theta(x)+q_1(x,0)\Big].
\label{annn_mu}
\end{align}
Note that to arrive at the last step in \eqref{annn_mu}, first four terms get cancelled with each other. Putting $q_1(x,0)$ (in Eq.~(18) in the \textit{Letter}) in terms of $p_0(x,0)$  back into above Eq. \eqref{annn_mu}, one gets the form given in Eq.~(19) in the \textit{Letter}.

Then using $p_0(x,0)$ in \eqref{1} and Eq. \eqref{2} into Eq.~(19) in the \textit{Letter}, the annealed CGF in dense limit reads,  
\begin{equation}
\mu_{\mathcal{A}}(\lambda)\simeq -\frac{\sqrt{4 D(1) T}}{\alpha}\; R_{\mathcal{A}}(0,\alpha\lambda\vert 1-\rho_a,1-\rho_b)
\end{equation}
where $R_{\mathcal{A}}$ is given by the Eq.~(20) in the \textit{Letter}.

\subsubsection*{Large deviation function}
\paragraph{}
Similar to quenched LDF, in this case the large deviation function is given by $\phi_\mathcal{A}(y)=-by-R_{\mathcal{A}}(0,b)$ where $b$ is determined in terms of $y$ by the condition, $\partial_b \phi_\mathcal{A}(y,b)=0$ (reported in Eq.~(26) in the \textit{Letter}). This condition implies,
\begin{equation}
    y=\frac{1}{2\sqrt{\pi}}\left[\big(1-\rho_b\big)e^{-b}-\big(1-\rho_a\big)e^{b}\right].
    \label{op_dense}
\end{equation}
Solving Eq. \eqref{op_dense} we find 
\begin{equation}
    b=-\log\big[u(y)\big] \quad \text{with}\quad  u(y)=\frac{y\sqrt{\pi}+\sqrt{\pi y^2+(1-\rho_a)(1-\rho_b)}}{1-\rho_b}.
\end{equation} Substituting $b$ back to $\phi_\mathcal{A}(y)$ we obtain a closed form of annealed LDF in dense case, which reads
\begin{equation}
    \phi_\mathcal{A}(y)=y~\log\big[u(y)\big]
    -\frac{1}{2\sqrt{\pi}}\left[(1-\rho_b)\Big(u(y)-1\Big)+(1-\rho_a)\left(\frac{1-u(y)}{u(y)}\right)\right].
    \label{Anne_dense_ldf}
\end{equation}

\section{Dilute limit}
\paragraph{}
In this case the calculations are similar to that of the dense case with important differences that $\lambda$ is of order of density and expansion of $Y$ contains non-vanishing $Y_0$. These lead to some technically different steps which are presented here.

\subsection{Quenched case}
\paragraph{}
Similar to the dense quenched case, expanding \eqref{qu_cgf}, quenched CGF in the dilute limit reads in the leading order
\begin{equation}
    \mu_\mathcal{Q}(\lambda)\simeq \lambda Y_0 - \frac{\sigma'(0)}{2}\int_{-\infty}^{\infty}dx~q_1~(\partial_xp_0)^2,
\end{equation}
which after using similar identity (16) in the \textit{Letter}, (just replacing $\sigma'(1)$ by $\sigma'(0)$ and $D(1)$ by $D(0)$) straightforwardly leads to Eq.~(21) in the \textit{Letter}. Now inserting the boundary conditions $q_1(x,0)$ and $p_0(x,T)$ along with the single-file constraint \eqref{Y0} into Eq.~(21)
of the \textit{Letter}, one arrives at
\begin{equation}
    \mu_\mathcal{Q}(\lambda)\simeq\lambda Y_0 + \rho_b \int_{0}^{\infty}dx~ \Big(p_0(x,0)- B\Big) 
    + \rho_a \int_{0}^{\infty}dx~ p_0(-x,0).
    \label{mucudil}
\end{equation}

Using $p_0(x,T)=B\,\Theta(x-Y_0)$ into (11a) in the \textit{Letter} we obtain, 
\begin{equation}
    p_0(x,0)=\frac{1}{\alpha}\log\left[1+\frac{e^{\alpha B}-1}{2}\text{erfc}\left(\frac{Y_0-x}{\sqrt{4D(0)T}}\right)\right].
    \label{3}
\end{equation}
Putting \eqref{3} into Eq. \eqref{mucudil} we obtain the quenched CGF in dilute case as,
\begin{align}
    \frac{\mu_{\mathcal{Q}}(\lambda)}{\sqrt{4D(0)T}}&\simeq\lambda~y+\frac{\rho_b}{\alpha}\int_{0}^{\infty}d\xi~\log\left[1+\frac{e^{-\alpha B}-1}{2}\text{erfc}(\xi-y)\right] + \frac{\rho_a}{\alpha}\int_{0}^{\infty}d\xi~\log\left[1+\frac{e^{\alpha B}-1}{2}\text{erfc}(\xi+y)\right]\cr
    &=\lambda~y+\frac{\rho_b}{\alpha}\int_{-y}^{\infty}d\xi~\log\left[1+\frac{e^{-\alpha B}-1}{2}\text{erfc}(\xi)\right] + \frac{\rho_a}{\alpha}\int_{y}^{\infty}d\xi~\log\left[1+\frac{e^{\alpha B}-1}{2}\text{erfc}(\xi)\right],
 \label{quenched_cgf_form_0}
\end{align}
where we define $y=\frac{Y_0}{\sqrt{4D(0)T}}$ ~and ~ $\xi=\frac{x}{\sqrt{4D(0)T}}$. Note that \eqref{quenched_cgf_form_0} is equivalent to Eq.~(22a) and (17b) in the \textit{Letter} where we define $b=\alpha B$. Note that in the dilute case the expression of $\mu_\mathcal{Q}(\lambda)$ is complete with the condition (22b) in the \textit{Letter}, derivation of which is presented below.

 The singularity of $q_1(x,t)$ at $(Y_0,T)$ in the dilute case means that the parameter $B$ cannot be determined self-consistently. Instead the parameter $B$ is determined from an optimization condition $\frac{d\mu_\mathcal{Q}}{dB}=0$ (refer \cite{Krapivsky2015tagged} for details) which leads to 
\begin{equation}
    \big(\alpha\lambda+\partial_y R_\mathcal{Q}\big)\frac{dy}{db}-\partial_b R_\mathcal{Q}=0,
    \label{7}
\end{equation}
where we denote $b=\alpha B$ and $R_\mathcal{Q}$ is defined in Eq.~(17b) of the \textit{Letter}.
We shall show that $\partial_b R_\mathcal{Q}=0$ using the relation
\begin{equation}
    \int_{Y_0}^{\infty}dx~q_1(x,T)=\int_{0}^{\infty}dx~q_1(x,0).
    \label{Y0}
\end{equation}
Then \eqref{7} implies that
$\partial_y R_\mathcal{Q}=-\alpha\lambda$. We conclude that Eq. \eqref{quenched_cgf_form_0} gives the quenched CGF where $b$ and $y$ are determined parametrically in terms of $\lambda$ from $\partial_b R_\mathcal{Q}=0$ and $\partial_y R_\mathcal{Q}=-\alpha\lambda$ respectively which are expressions equivalent to Eq.~(22a) and (22b) in the \textit{Letter}.

To prove that indeed $\partial_b R_\mathcal{Q}=0$, we use the solution of $q_1(x,T)$, obtained from Eq.~(11b) in the \textit{Letter}, and the quenched initial condition $q_1(x,0)$ into \eqref{Y0} and find, 
\begin{align}
    \int_{0}^{\infty}dx~\rho_b=e^{\alpha B}\int_{-\infty}^{\infty}dz~\frac{q_1(z,0)A(Y_0-z)}{1+(e^{\alpha B}-1)A(Y_0-z)},
    \label{4}
\end{align}
where we define
\begin{equation*}
    A(Y_0-z)=\int_{Y_0}^{\infty}dx~\frac{1}{\sqrt{4\pi D(0)T}}\exp\left(-\frac{(z-x)^2}{4 D(0)T} \right)=\frac{1}{2}\text{erfc}\left(\frac{Y_0-z}{\sqrt{4D(0)T}}\right).
\end{equation*}
Eq. \eqref{4} can be re-written as 
\begin{align}
    \rho_a\int_{-\infty}^{0}dz~\frac{e^{\alpha B}A(Y_0-z)}{1+(e^{\alpha B}-1)A(Y_0-z)}=\rho_b\int_{0}^{\infty}dz~\left[1-\frac{e^{\alpha B}A(Y_0-z)}{1+(e^{\alpha B}-1)A(Y_0-z)}\right].
    \label{5}
\end{align}
Now changing variable $x=Y_0-z$ in the left side and $x=z-Y_0$ in the right side of the \eqref{5} we obtain 
\begin{align}
    \rho_a\int_{Y_0}^{\infty}dx~\frac{e^{\alpha B}A(x)}{1+(e^{\alpha B}-1)A(x)}&=\rho_b\int_{-Y_0}^{\infty}dx~\left[1-\frac{e^{\alpha B}A(-x)}{1+(e^{\alpha B}-1)\Big(1-A(x)\Big)}\right]\cr
    &=\rho_b\int_{-Y_0}^{\infty}dx~\left[1-\frac{A(-x)}{\Big\{1+(e^{-\alpha B}-1)A(x)\Big\}}\right]\cr
    &=\rho_b\int_{-Y_0}^{\infty}dx~\frac{e^{-\alpha B}A(x)}{1+(e^{-\alpha B}-1)A(x)}
    \label{6}
\end{align}
where we use $A(x)+A(-x)=1$, implied from the identity, $\text{erfc}(x)+\text{erfc}(x)=2$. This means $Y_0$ is such that it follows the relation in Eq.~\eqref{6}.

In terms of re-scaled variables we get an identity from \eqref{6},
\begin{equation}
    \rho_a\int_{y}^\infty d\xi\frac{e^{b}\frac{1}{2}\text{erfc}(\xi)}{1 + (e^{b}-1)\frac{1}{2}\text{erfc}(\xi)}
    =\rho_b\int_{-y}^\infty d\xi\frac{e^{-b}\frac{1}{2}\text{erfc}(\xi)}{1 + (e^{-b}-1)\frac{1}{2}\text{erfc}(\xi)}.
    \label{iden_dil_qu}
\end{equation}
From the above identity \eqref{iden_dil_qu} it is straight-forward to verify $\partial_b R_\mathcal{Q}=0$ where $R_\mathcal{Q}$ is defined in Eq.~(17b) in the \textit{Letter} for the dilute case. 

\subsection{Annealed case}
\paragraph{}
Here we present the steps to derive Eq.~(24) starting from Eq.~(23) in the \textit{Letter} and finally obtain the CGF in \eqref{an_cgf} in the leading order.
Similar to the dense annealed case, we expand $F[q(x,0)]$ in leading order and find
\begin{align*}
    F[q(x,0)]&\simeq\frac{1}{\alpha}\int_{-\infty}^{\infty}dx~\int_{\Bar{\rho}(x)}^{q_1(x,0)}dz~\left(\frac{q_1(x,0)}{z}-1\right)\\
    &=\frac{1}{\alpha}\int_{-\infty}^{\infty}dx~\left[q_1(x,0)\log\bigg(\frac{q_1(x,0)}{\Bar{\rho}(x)}\bigg)+\Big\{\Bar{\rho}(x)-q_1(x,0)\Big\}\right].
\end{align*}
Then using Eq.~(23) in the \textit{Letter} in \eqref{an_cgf} we find the annealed CGF in leading order as
\begin{align} \mu_{\mathcal{A}}(\lambda)&\simeq\lambda~Y_0+\int_{-\infty}^{\infty}dx~q_1(x,0)p_0(x,0)-\int_{-\infty}^{\infty}dx~q_1(x,T)p_0(x,T)\cr
&~~~-\int_{-\infty}^{\infty}dx~q_1(x,0)~\Big(p_0(x,0)-B\Theta(x)\Big) + \frac{1}{\alpha}\int_{-\infty}^{\infty}dx~\Big[q_1(x,0)-\Bar{\rho}(x)\Big]\cr
&=\lambda Y_0+\frac{1}{\alpha}\int_{-\infty}^{\infty}dx~\Big[q_1(x,0)-\rho_a \Theta(-x) -\rho_b \Theta(x)\Big],
\label{annmu}
\end{align}
where we incorporate $p_0(x,T)$ and single-file constraint in \eqref{Y0} and then it is straightforward to note the cancellation of all terms except the first and last one. Note that \eqref{annmu} is equivalent to Eq.~(24) in the \textit{Letter}. Finally, the annealed CGF in this limit is given by, 
\begin{align*}
    \mu_{\mathcal{A}}(\lambda)\simeq \lambda Y_0+ \frac{\rho_a}{\alpha}~ \int_{0}^{\infty}dx~\Big(e^{\alpha p_0(-x,0)}-1\Big) +\frac{\rho_b}{\alpha}~ \int_{0}^{\infty}dx~\Big(e^{-\alpha B+\alpha p_0(x,0)}-1\Big).
\end{align*}

Now using \eqref{3} one gets,
\begin{align}
    \frac{\mu_{\mathcal{A}}(\lambda)}{\sqrt{4D(0)T}}&\simeq \lambda y+\frac{1}{\alpha}\left[\rho_b\int_{0}^{\infty}d\xi~\frac{e^{-\alpha B}-1}{2}\text{erfc}(\xi-y)+\rho_a\int_{0}^{\infty}d\xi~\frac{e^{\alpha B}-1}{2}\text{erfc}(\xi+y)\right]\cr
    &= \lambda y+\frac{1}{\alpha}\left[\rho_b\int_{-y}^{\infty}d\xi~\frac{e^{-\alpha B}-1}{2}\text{erfc}(\eta)+\rho_a\int_{y}^{\infty}d\xi~\frac{e^{\alpha B}-1}{2}\text{erfc}(\eta)\right]\cr
    &= \frac{1}{\alpha}\; \Big\{\alpha\lambda y + R_{\mathcal{A}}(y,b\vert \rho_a,\rho_b)\Big\}
    \label{qu_dil_ann}
\end{align}
where $R_\mathcal{A}$ is given by Eq.~(20)
in the \textit{Letter} with $b=\alpha B$. Eq.~\eqref{qu_dil_ann} is the result for the CGF in the annealed case reported in Eq.~(22a) with the subscript $\mathcal{Q}$ replaced by $\mathcal{A}$.

The variables $y$ and $b$ are determined from additional conditions  $\partial_y R_\mathcal{A}=-\alpha\lambda$ and $\partial_b R_\mathcal{A}=0$ (equivalent to Eq.~(22b) in the \textit{Letter} with the subscript $\mathcal{Q}$ replaced by $\mathcal{A}$). These two relations come from the single-file constraint in \eqref{Y0} for the annealed case and the optimization condition $\frac{d\mu_\mathcal{A}}{dB}=0$ (refer \cite{Krapivsky2015tagged} for details). Eq. \eqref{Y0} gives the condition $ \partial_b R_\mathcal{A}=0$ as we show below.
Using $p_0(x,0)$ from \eqref{3} in \eqref{Y0} we get,
\begin{align}
    \rho_a e^b \int_{-\infty}^{0}d\xi~\frac{1}{2}&\text{erfc}\left(y-\xi\right)+\rho_b  \int_{0}^{\infty}d\xi~\frac{1}{2}\text{erfc}\left(y-\xi\right)\cr
   & =\rho_b e^{-b} \int_{0}^{\infty}d\xi~\left[1+\frac{e^b-1}{2}\text{erfc}(y-\xi)\right],
   \label{random}
\end{align}
where $b=\alpha B$ and the $y= \frac{Y_0}{\sqrt{4\pi D(0)T}}$. Subsequently using the relation $\text{erfc}(\xi)=2-\text{erfc}(-\xi)$ in the right hand side of Eq. \eqref{random} followed by a variable change leads to an identity, 
\begin{align}
    \rho_a e^{b}\int_{y}^\infty d\xi~\frac{1}{2}\text{erfc}(\xi) = \rho_b e^{-b}\int_{-y}^\infty d\xi~\frac{1}{2}\text{erfc}(\xi).
    \label{iden_ann}
\end{align}
 Using the above identity it is straight-forward to show $\partial_b R_\mathcal{A}=0$ where $R_\mathcal{A}$ is defined in Eq.~(20) in the \textit{Letter}. 
 
 The additional condition $\frac{d\mu_\mathcal{A}}{dB}=0$ with $\mu_\mathcal{A}$ defined in \eqref{qu_dil_ann} gives
\begin{equation}
    \big(\alpha\lambda+\partial_y R_\mathcal{A}\big)\frac{dy}{db}-\partial_b R_\mathcal{A}=0,
    \label{10}
\end{equation}
which using $\partial_b R_\mathcal{A}=0$ leads to the relation $\partial_y R_\mathcal{A}=-\alpha\lambda$.
\subsubsection*{Large deviation function}
\paragraph{}
From the definition of Large deviation function in (25) of the \textit{Letter}, the annealed LDF in dilute case is given by
\begin{equation}
    \phi_\mathcal{A}(y)=-R_\mathcal{A}(b,y)=\rho_a\Big(1-e^{b}\Big)h(y)  
    +\rho_b\Big(1-e^{-b}\Big)h(-y)
    \label{phiA}
\end{equation}
where $h(y)=\int_{y}^\infty d\eta~\frac{1}{2}\text{erfc}(\eta)$ and $b$ is determined from $\partial_b R_\mathcal{A}=0$ which is equivalent to Eq. \eqref{iden_ann}. Using the solution of $b$ from \eqref{iden_ann} into the Eq. \eqref{phiA} we find a compact form of the LDF in the annealed dilute case,
\begin{align}
    \phi_\mathcal{A}(y)&= \rho_a h(y) +\rho_b h(-y)-2\sqrt{\rho_a\rho_b h(y)h(-y)}\cr
    &=\left(\sqrt{\rho_a h(y)}-\sqrt{\rho_b h(-y)}\right)^2.
\end{align}

\section{Relation of $\alpha$ with $K_T$}
The fluctuation dissipation relation relates \cite{Derrida2009,Krapivsky2015PRL,Krapivsky2015tagged} the diffusivity $D(\rho)$ and the mobility $\sigma(\rho)$ by
\begin{equation}
    \frac{2D(\rho)}{\sigma(\rho)}=f''(\rho)
\end{equation}
where $f(\rho)$ is the canonical free energy density (inverse temperature $\beta$ is set to value one). Similarly, the isothermal compressibility $K_T$ relates to the free energy density by 
\begin{equation}
    K_T=\frac{1}{\rho^2 f''(\rho)}.
\end{equation}
From the two relations,
\begin{equation}
    \frac{2D(\rho)}{\sigma(\rho)}=\frac{1}{\rho^2K_T}.
\end{equation}
Considering the perturbation expansion in the low density  limit, $D(\rho)\simeq D(0)$ and $\sigma(\rho)\simeq \rho \sigma'(0)$, gives
\begin{equation}
    \rho K_T\simeq \frac{\sigma'(0)}{2D(0)}\equiv\alpha.
\end{equation}
 Similarly, in the dense limit, considering $D(\rho)\simeq D(1)$ and $\sigma(\rho)\simeq (\rho-1)\sigma'(1)$, we get
\begin{equation}
    \frac{K_T}{1-\rho}\simeq -\frac{\sigma'(1)}{2D(1)}\equiv -\alpha.
\end{equation}

\bibliography{references}
\bibliographystyle{ieeetr}